\documentclass[twocolumn,preprintnumbers,
endnote,nofootinbib,prl]{revtex4}
\usepackage{graphicx}

\usepackage[hypertex]{hyperref}
\newcommand{\vev}[1]{\langle {#1} \rangle}
\newcommand{\lsim}{\lesssim}
\newcommand{\gsim}{\gtrsim}

\newcommand{\eq}[1]{Eq.~(\ref{#1})}

\newcommand{\ord}[1]{\mathcal{O}{(#1)}}
\newcommand{\beq}{\begin{equation}}
\newcommand{\eeq}{\end{equation}}
\newcommand{\eps}{\varepsilon}
\newcommand{\mum}{\mu_-}
\newcommand{\mup}{\mu_+}

\begin{document}

\pagestyle{plain}

\title{Nucleon Decay into Dark Sector}

\author{Hooman Davoudiasl\footnote{email: hooman@bnl.gov}
}
\affiliation{Department of Physics, Brookhaven National Laboratory,
Upton, NY 11973, USA}


\begin{abstract}
 
A sub-GeV dark sector fermion $X$ can have baryon number violating interactions induced 
by high scale physics, leading to nucleon decay into $X + \text{meson}$ and $\text{neutron}\to X + \text{photon}$.  
Such processes can mimic standard search modes containing a neutrino, but have different kinematics and
may have escaped detection.  If a dark force mediated by a light vector $Z_d$ acts on $X$, depending on parameters,   
$\text{neutron} \to X + Z_d$ can be important.  In typical scenarios, $Z_d$  
decays into $\ell^+\ell^-$, where $\ell=e,\mu$, with order unity branching fraction.  
Nucleon decay searches can potentially uncover new dark states 
that are otherwise inaccessible, due to their negligible coupling to ordinary matter or cosmological abundance.

\end{abstract}
\maketitle

What constitutes dark matter (DM) remains one of the open questions
of particle physics and cosmology.  In broad terms, DM can only have  
feeble interactions with visible matter, but its possible mass covers a 
wide range, from sub-eV to well above the TeV scale.
However, for example, if the similar baryon and DM energy densities 
are due to asymmetries from a common underlying mechanism \cite{adm,Kaplan:2009ag}, GeV-scale
DM is motivated; for some recent reviews of these scenarios see Ref.~\cite{reviews}.
Regardless of its origin, once the mass of DM is near or below $\sim 1$~GeV,
detecting it directly via scattering from target atoms, characterized
by soft recoils, poses a significant challenge.  Hence, it
is worthwhile to examine alternative probes of DM with
mass $\lsim 1$~GeV \cite{Kile:2009nn,lightDM,Essig:2011nj,Essig:2012yx}.  We also
note that in many models  the ``dark" sector includes other 
states or new forces \cite{Pospelov:2007mp,ArkaniHamed:2008qn,Essig:2013lka} 
that lack any direct interaction with the Standard Model (SM) and can be quite light.    

In this work, we introduce a dark fermion $X$ whose mass $m_X$
is below the nucleon mass $m_N \simeq 0.94$~GeV.  
For now, we will assume that $X$ is a singlet; we will discuss the possibility of $X$ having 
dark gauge charges later.  
Further, we will assume that $X$ has baryon number
violating interactions suppressed by a high scale $M\gg 1$~TeV and dominated by the 
``neutron portal" \cite{hylogenesis} operator 
\beq
O_{\rm BV} = \frac{(X u d d)_R}{M^2},
\label{OBV}
\eeq
where $u$ and $d$ denote the up and down quarks, respectively; the subscript $R$
refers to the right-handed chirality of the fields.   Other
operators can be considered, but our choice suffices to
demonstrate the main relevant features.  The above type of interaction may arise
at high scales on general grounds and has been employed in a variety of
models \cite{Shelton:2010ta,hylogenesis,Davoudiasl:2013pda}.  
In our scenario, a nucleon $N$ could potentially decay via 
$N \to X +$ meson \cite{Davoudiasl:2013pda}, or $N\to X + \text{photon}$.  For certain values of $M$, 
such decays may be accessible at nucleon decay experiments.  Those experiments, however, 
are largely motivated by grand unified models \cite{Georgi:1974sy,Pati:1974yy}, with signals that 
typically have different characteristics from the dark sector signals discussed here.  Thus, the current bounds 
do not directly apply to our setup, though they provide general guidance.  In particular, it is conceivable that 
some of the signals discussed below might go undetected in existing analyses.   

Before presenting the details, we note that much of our discussion will be
relevant to any sufficiently long-lived dark or weakly-interacting 
sector fermion that couples to baryons via \eq{OBV}.
Therefore, $X$ does not have to constitute cosmic DM, though that is a distinct possibility.
Hence, like accelerator-based probes, and in contrast to typical direct or indirect searches for DM,
our nucleon decay signal does not depend on the local abundance of $X$.  However, whereas
accelerator-based experiments require that $X$ couple to ordinary matter with at least some modest strength,
the dark states considered here can have negligible interactions.
For example, $X_R$ could also be a sub-GeV ``right-handed neutrino"
associated with a seesaw mechanism for small neutrino masses \cite{seesaw}.  Such states can possibly 
play a role in cosmology \cite{Asaka:2005an} and may  
lead to interesting phenomenology \cite{de Gouvea:2007uz,Atre:2009rg}.

The operator $O_{\rm BV}$ in \eq{OBV} leads to interactions of $X$ with baryons and mesons, at
low energies.  These interactions can be studied using chiral perturbation 
theory \cite{Claudson:1981gh}, which would yield fair estimates
for meson momenta $\ord{100~\text{MeV}}$, typical of the regime of interest here.   
Using chiral perturbation formalism \cite{Claudson:1981gh},
we obtain the following baryon number preserving, $\Delta B = 0$,
interactions of the proton $p$ and the neutron $n$ with pions and the $\eta$
\begin{eqnarray}
{\cal L}_0 &=&
\frac{D+F}{2 f_\pi}
\left[\left(\frac{3F-D}{D+F}\right)\frac{\partial_\mu \eta}{\sqrt{3}}-\partial_\mu \pi^0 \right]
\bar n \,\gamma^\mu \gamma_5 \, n \nonumber\\
&+& \left(\frac{D+F}{\sqrt{2} f_\pi}\right) \partial_\mu \pi^-\, \bar n \,\gamma^\mu \gamma_5 \, p + \ldots\,.
\label{L0}
\end{eqnarray}
Baryon number violating $\Delta B =1$ interactions are given by \cite{Claudson:1981gh} 
\begin{eqnarray}
{\cal L}_1 =\beta \, c_1  \, \overline{X^c}\left\{n_R - \frac{i}{f_\pi} \left[\frac{p_R}{\sqrt{2}}\pi^- 
+ \frac{n_R}{2} \left(\sqrt{3} \eta - \pi^0\right)\right]\right\},  
\label{L1}
\end{eqnarray}
where $D=0.80$, $F=0.47$, $\beta = 0.012 \pm 0.0026$~GeV$^3$ \cite{Aoki:2008ku}, and $f_\pi \simeq 92.2$~MeV.
Here, $c_1 \equiv 1/M^2$ sets the strength of baryon number violation in \eq{OBV}; $p_R$ and
$n_R$ denote right-handed projections of the proton and neutron, respectively (see also Ref.~\cite{Davoudiasl:2011fj}).  
In Eqs.(\ref{L0}) and (\ref{L1}), we have left out terms that are not directly relevant to our discussion.  
In \eq{L1}, the first term describes the mixing of the neutron and $X$ through the mass mixing parameter
$\beta c_1 \ll m_N$, whose effect on the spectrum can be safely ignored.

We first examine the case where the mesons are emitted on-shell in $N \to X + \text{meson}$: 
$p\to X \pi^+$ and $n\to X \pi^0$, $X \eta$.  If $m_X \ll m_N-m_{\text{meson}}$,
then these decays will effectively look like the standard nucleon decay $N \to \text{neutrino + meson}$,
and it will be difficult to infer the effect of the interaction in \eq{OBV}.  Hence, we will be mostly
interested in the range $\text{few}\times 100~\text{MeV}\lsim m_X \lsim m_N$.

There are two contributions to $N \to X + \text{meson}$.  One contribution arises from $\Delta B = 0$ interactions
in ${\cal L}_0$, with the outgoing nucleon mixing into $X$, via the first term in ${\cal L}_1$.
The other originates from $\Delta B=1$ couplings of nucleons
to mesons and $X$ in ${\cal L}_1$. We find the rates
\begin{eqnarray}
&\Gamma& \!\!(p\to X \pi^+) = \frac{\beta^2 c_1^2 \, |\vec{p}_{\pi^+}|}{32 \pi f_\pi^2 m_p^2}\label{Gamp} \\
&\times& \left[(A_p^2 + B_p^2)f(m_p, m_{\pi^+})+(B_p^2-A_p^2)m_p m_X\right],\nonumber
\end{eqnarray}
where $f(x,y)\equiv (x^2 - y^2 + m_X^2)/2$,
\beq
A_p = 1 + \frac{m_p + m_X}{m_p-m_X}(D+F)\,,
\label{Ap}
\eeq
and $B_p$ can be obtained from $A_p$ by $m_X\to -m_X$.
We also find
\begin{eqnarray}
&\Gamma& \!\!(n\to X \phi) = \frac{k_\phi \beta^2 c_1^2 \, |\vec{p}_\phi|}{64 \pi f_\pi^2 m_n^2}\label{Gamn} \\
&\times& \left[(A_{n\phi}^2 + B_{n\phi}^2)f(m_n, m_\phi)+(B_{n\phi}^2 - A_{n\phi}^2)m_n m_X\right],\nonumber
\end{eqnarray}
where $\phi=\pi^0, \eta$ and $k_\phi = 1(3)$ for $\pi^0 (\eta)$.  The expressions for $(A_{n\pi^0}, B_{n\pi^0})$
are obtained from those of $A_p$ and $B_p$ with $p\to n$, and $(A_{n\eta}, B_{n\eta})$ can in turn be
obtained from $(A_{n\pi^0}, B_{n\pi^0})$ with the substitution $D+F \to (3F-D)/3$.

With our assumptions, we can then approximate the total width of the proton by $\Gamma (p\to X \pi^+)$.  The
width of the neutron, bound in a nucleus, resulting from \eq{OBV} is to a good approximation given by
$\Gamma_n = \Gamma (n\to X \pi^0) + \theta (m_n - m_X - m_\eta) \Gamma (n\to X \eta)$.  
\begin{figure}[t]
\includegraphics[width=0.46\textwidth,clip]{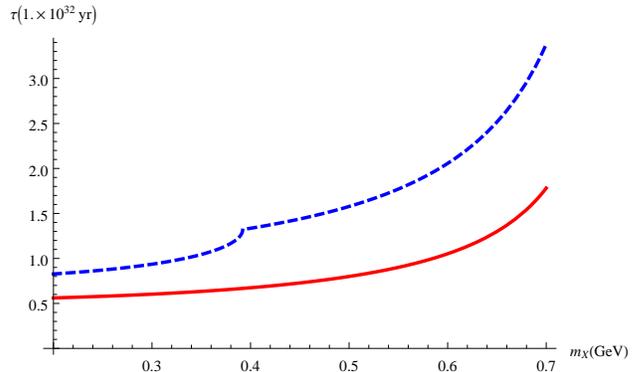}
\caption{The lifetimes of nucleons $p$ (solid) and $n$ (dashed) as a function of the dark fermion mass $m_X$, assuming
decay into on-shell mesons, via the interaction in \eq{OBV} with $c_1^{-1/2}=M=10^{15}$~GeV (for more details, see the text).}
\label{fig:lifetimes}
\end{figure}

In Fig.\ref{fig:lifetimes},
we have presented the resulting lifetimes of $p$ (solid) and $n$ (dashed), 
for $c_1^{-1/2}=M=10^{15}$~GeV, as a function of $m_X$.
The discontinuity in the neutron lifetime curve signifies the $\theta$-function treatment of the $\eta$ threshold in
the two-body decay.  The typical nucleon lifetime $\tau$ comes out to be $\sim 10^{32}$~yr,
which is not far from the current bounds on nucleon decay
into meson + missing energy: $\tau(p\to \pi^+ \bar \nu) > 1.6\times 10^{31}$~yr \cite{Wall:2000pq}
and $\tau[n\to \pi^0 (\eta)  \bar \nu] > 1.12 (1.58)\times 10^{32}$~yr \cite{McGrew:1999nd}.  
Those bounds, however, are obtained under the assumption that the missing energy
comes from neutrinos and are subject to specific kinematic cuts and selection criteria.  Note that in
our setup $m_X$ does not need to be small and as a result the meson can come out with low momentum, which
can lead to missed events.  For instance, in the Soudan 2 experiment, 
the pion momentum in $p \to \pi^+ \bar \nu$
was required to be between 140 and 420~MeV \cite{Wall:2000pq}.  
The pion momentum for this process is about 459~MeV,
but simulations suggest that pions emerging from within the iron nucleus, on average,
lose about half their momentum \cite{Wall:2000pq}.  While the standard search cut 
can largely accommodate the momentum loss in the nuclear medium, for $p \to X \pi^+$ in our model the initial pion
momentum can be small enough that it may fall outside the range chosen by the experiment.
For example, with $m_X=600$~MeV, initially we have $p_{\pi^+}\simeq 251$~MeV,
half of which would be too small to pass the above experimental cut on momentum.

Due to its magnetic dipole moment, the neutron can 
interact with a photon $\gamma$ \cite{KS}, which in our scenario can lead to 
$n\to X \gamma$.  For a photon of momentum $q$, the dipole interaction is given by  
\beq
\frac{i \, e}{2 m_p} \bar n \, \sigma^{\mu\nu}q_\nu F_2(q^2) \, n \, A_\mu\,,
\label{dipole}
\eeq
where $e \equiv \sqrt{4\pi \alpha}$ is the electromagnetic 
coupling constant and $F_2(q^2)$ is a form factor.  In this work the photon is on-shell, with $q^2=0$, and  
$F_2(q^2) = F_2(0) \simeq -1.91$ \cite{pdg}.  We find the decay rate 
\beq
\Gamma(n\to X \gamma) = \frac{\alpha \, \beta^2 c_1^2 \, F_2(0)^2}{16 \,m_p^2\, m_n^3} (m_n^4 - m_X^4)\,.
\label{GamnXgam}
\eeq
Note that, compared to the rates in Eqs.~(\ref{Gamp}) and (\ref{Gamn}), 
the above rate is suppressed by $\sim 4 \pi \alpha \, (f_\pi/m_p)^2 \sim 10^{-3}$.  For example, 
with $c_1^{-1/2}=10^{15}$~GeV, as in Fig.\ref{fig:lifetimes}, and $m_X = 700$~MeV, 
we obtain $\tau(n\to X \gamma) \simeq 1.2 \times 10^{35}$~yr and   
nucleon decay will typically be dominated by $X+\text{meson}$ final states.  

For $m_N-m_X<m_\pi$, three-body nucleon decays through 
off-shell pion  states $\pi^*$ are quite suppressed and we find that the leading 
nucleon decay channel will be $n\to X \, \gamma$.  
We present the rate for $n\to X\, \pi^{0*}\to X \gamma \gamma$ 
in the appendix, as it constitutes  
the main off-shell meson channel.  To see this, note that whereas
the neutral pion $\pi^0$ decays electromagnetically at 1-loop, the charged
pion $\pi^+$ (from proton decay) can only decay through 
weak interactions suppressed by the heavy $W$ mass.   
For $c_1^{-1/2}=10^{15}$~GeV and $m_X=840$~MeV, we find
$\tau(n\to X \gamma\gamma) \simeq 7.1 \times 10^{41}$~yr, from \eq{ratenX2a}, 
whereas $\tau(n\to X \gamma) \simeq 2.2 \times 10^{35}$~yr, from \eq{GamnXgam}, well above 
the current bound on $\tau(n\to \nu \gamma) > 2.8 \times 10^{31}$~yr \cite{pdg}.  
If pions cannot be emitted on-shell and $n\to X \, \gamma$ is the main nucleon decay channel, 
$c_1^{-1/2} \gsim 10^{14}$~GeV could be allowed by current bounds.  Note also that 
the photon energy here is typically softer than in $n\to \nu \gamma$, which 
can affect the efficiency of the search.

\underline{\it The Effect of Dark Forces:}
So far, we have ignored the dark sector interactions of $X$.  However, for a variety
of reasons, one may expect the presence of dark forces that act on $X$.  For example,
if $X$ is asymmetric DM there needs to be a mechanism to annihilate
$(X, \bar X)$ pairs efficiently.  This can be achieved in a minimal fashion by introducing a
gauged $U(1)_d$, mediated by a vector boson 
$Z_d$ of mass $m_{Z_d}<m_X$ \cite{hylogenesis} that couples to $X$.  
The vector $Z_d$ is assumed to interact with the visible sector weakly, for example
through kinetic mixing \cite{Holdom:1985ag} of $U(1)_d$ with $U(1)_Y$ hypercharge,
parametrized by $\varepsilon\ll 1$, which will be assumed in what follows  
(for simplicity, we will ignore possible $Z$-$Z_d$ 
mass-mixing effects \cite{Davoudiasl:2012ag,Davoudiasl:2012qa,Davoudiasl:2013aya}).  
Such light vector bosons have also been invoked \cite{directg-2mu,Pospelov:2008zw} as a possible
explanation for the deviation  
of the measured muon $g-2$ from the SM expectation \cite{pdg} 
and are subject of various experimental 
searches \cite{Essig:2013lka}.     

Let us examine the consequences of introducing a $U(1)_d$ charge for $X$.     
For $O_{\text{BV}}$ in \eq{OBV} to be gauge invariant, we assume that
only the $X_L$ chirality of $X$ has charge $Q_d$ under $U(1)_d$.
This leads to a new $n X Z_d$ interaction in our model through $n$-$X$
mixing in \eq{L1}, and provides a mechanism for $n\to X Z_d$.  
We will mainly consider the case with $m_n-m_X >m_{Z_d}$, since 
decay through off-shell $Z_d$ is significantly suppressed by $\alpha \eps^2$ 
(see the appendix).  Assuming that there are no  
$U(1)_d$ charged states of mass $<m_{Z_d}/2$, 
kinetic mixing typically yields $\ord{1}$ branching 
fractions for $Z_d \to \ell^+\ell^-$, with $\ell=e,\mu$.  
Treating $n$-$X$ mixing as a mass insertion, we obtain 
the spin-averaged squared amplitude for $n\to X Z_d$ 
\beq
2\left(\frac{Q_d \,g_d\, \beta c_1  m_X}{m_n^2 - m_X^2}\right)^2
\left[\frac{p_X.p_n}{2} + \frac{(p_n.p_{Z_d})(p_X.p_{Z_d})}{m_{Z_d}^2}\right],
\label{amp2}
\eeq
where $g_d$ is the $U(1)_d$ coupling constant and 
the 4-momenta are labeled in obvious notation in the scalar products.
Ignoring terms of $\ord{m_{Z_d}^2}$, the associated decay rate can be written as
\beq
\Gamma (n\to X Z_d) \simeq \frac{Q_d^2 \, \alpha_d}{8} \left(\frac{\beta \, c_1 \, m_X}{m_n^{3/2}}\right)^2
\left(\frac{\mup^2}{\mum^2} + \frac{\mum^2}{m_{Z_d}^2}\right),
\label{GamntoXZd}
\eeq
where $\alpha_d \equiv g_d^2/(4 \pi)$ and $\mu_\pm^2\equiv m_n^2 \pm m_X^2$.

For $m_n - m_X > m_{\pi^0}, m_{Z_d}$, we typically find that
$n\to X \pi^0$ and $n\to X Z_d$ can give comparable rates.  
For example, if $Q_d=1$, $\alpha_d=\alpha$, $c_1^{-1/2} = 10^{15}$~GeV, 
$m_X = 700$~MeV, and $m_{Z_d}=50$~MeV, we have 
$\tau(n\to X \pi^0) \simeq 3.4\times 10^{32}$~yr while 
$\tau(n\to X Z_d) \simeq 1.7 \times 10^{33}$~yr, from \eq{Gamn} and
\eq{GamntoXZd}, respectively.

With $m_{\pi^0} > m_n - m_X > m_{Z_d}$, the $n\to X Z_d$ 
channel can be dominant and have a larger rate than $n\to X \gamma$.  For instance,
if we raise $m_X$ to 840~MeV in the preceding example, we find
$\tau(n\to X Z_d) \simeq 2.3\times 10^{33}$~yr and
$\tau(n\to X \gamma) \simeq 2.2 \times 10^{35}$~yr from \eq{GamnXgam}.
Here, the rate for $n\to X Z_d$ dominates over
$n\to X \gamma$, as long as $Q_d^2 \, \alpha_d \gsim 10^{-2} \alpha$.  
With $\ord{1}$ branching fraction for $Z_d\to \ell^+\ell^-$, the relevant bound is 
$\tau(n\to e^+ e^- \nu)>2.6\times 10^{32}$~yr \cite{pdg}, suggesting 
$c_1^{-1/2} \gsim 5 \times 10^{14}$~GeV for $Q_d^2 \alpha_d=\alpha$.  
However, in our case the signature is a distinct dilepton resonance at the $Z_d$ mass.  Note that 
if $Z_d$ decayed invisibly in this example, the dominant nucleon decay 
mode would be entirely invisible and subject to the bound $\tau(n\to 3 \,\nu) > 5\times 10^{26}$~yr \cite{pdg}  
which would imply $c_1^{-1/2} \gsim 2\times 10^{13}$~GeV. 

A few comments on the preceding discussion are in order.  First of all, the
second term in \eq{GamntoXZd} seems to diverge as $m_{Z_d}\to 0$.
However, in a physical theory this will not be the case.  For example, 
if the mass of $Z_d$ originates from the vacuum expectation value of a scalar $\phi$ 
with $U(1)_d$ charge $Q_d=1$ then $m_{Z_d} \sim g_d \vev{\phi}$.  Since $X_R$ in \eq{OBV} has no
dark charge but $X_L$ does, $m_X \propto \vev{\phi}$ and as $\vev{\phi}\to 0$ the second term in
\eq{GamntoXZd} is finite.  Alternatively, if $X_R$ were charged under $U(1)_d$, then
$c_1 \propto \vev{\phi}$ in order for $O_{\text{BV}}$ to be gauge invariant and 
terms that scale as $1/m_{Z_d}^2$
would again be finite as $\vev{\phi}\to 0$.  In either case, if $g_d\to 0$ 
only the term $\propto m_{Z_d}^{-2}$, associated with the Goldstone boson, survives.  

Secondly, if $X$ is a significant component of DM, then direct detection limits from Ref.~\cite{Essig:2012yx} constrain its
electron scattering cross section $\sigma_e < 10^{-37}$~cm$^2$, at 90 \% CL.  In that case, if the $Z_d$
couples to the SM through kinetic mixing \cite{Holdom:1985ag}, then the mixing parameter is constrained 
by $\varepsilon \lsim 10^{-3}$ \cite{Essig:2011nj} for $m_{Z_d}\sim 100$~MeV.  
We note that, for this range of $m_{Z_d}$ values, the decay length for $Z_d\to \ell^+\ell^-$
will be $\lsim 1$~m \cite{Davoudiasl:2012ag}, as long as $\varepsilon \gsim 10^{-6}$, 
and hence typically contained within the experimental fiducial volume.

To summarize, sub-GeV states $X$ from a dark sector, such
as dark matter or right-handed neutrinos, may have
baryon number violating couplings to the SM sector, suppressed by 
a high scale.  Nucleons could then decay into $X$
with dominant rates, potentially measurable at current or planned experiments.
These decays can have kinematical
features quite distinct from those of standard search processes containing a neutrino.  
The differences in kinematics also imply that
existing event selection criteria may not be sensitive to our signals, 
allowing looser bounds than those implied by current analyses.  We pointed out that if $X$
couples to a low mass dark force carrier $Z_d$ then a new
neutron decay mode $n \to X \,Z_d$ can emerge in our scenario and possibly dominate nucleon decays.  
If $Z_d$ kinetically mixes with the photon, one typically expects $Z_d$ to decay into charged leptons 
with $\ord{1}$ branching fraction.

In conclusion, nucleon decay can potentially provide an
interesting probe of dark sector states, even when they have tiny 
couplings to the visible sector or negligible cosmic abundance and other approaches are impractical.

\acknowledgments

We thank P. Huber, I. Lewis, W. Marciano, J. Millener, and K. Sigurdson for discussions.
This work is supported in part by the United States Department of Energy
under Grant Contracts DE-AC02-98CH10886.

\section{Appendix}
\appendix*

To calculate $n\to X\, \pi^{0*}\to X \gamma \gamma$, we need the $\pi^0\gamma\gamma$
coupling, given by the Wess-Zumino-Witten term \cite{WZW}
\beq
{\cal L}_{\pi^0 \gamma\gamma} = \frac{N_c \,\alpha}{24 \pi f_\pi}\pi^0
\varepsilon^{\mu\nu\lambda\sigma} F_{\mu\nu} F_{\lambda\sigma}\,,
\label{WZW}
\eeq
with $N_c=3$ the number of quark colors and $F_{\mu\nu}$ the electromagnetic
field strength tensor.  
Then, using Eqs. (\ref{L0}), (\ref{L1}), and (\ref{WZW}), we find the following
differential decay rate 
\begin{eqnarray}
\frac{d \Gamma (n\to X \gamma\gamma)}{d s_{\gamma\gamma}} &=&
\frac{\kappa^2 \sqrt{(\mum^2 + s_{\gamma\gamma})^2 - 4 m_n^2 s_{\gamma\gamma}}}
{2^{12} \pi^3 m_n^3} \nonumber \\
\times\frac{s_{\gamma\gamma}^2}{(s_{\gamma\gamma}-m_{\pi^0}^2)^2} && \hskip-0.6cm
\left[x_1 \, (\mup^2 - s_{\gamma\gamma}) + 2 x_2 \, m_n m_X\right]
\label{diffnX2a}
\end{eqnarray}
as a function of the di-photon invariant mass $s_{\gamma\gamma} \equiv m_{\gamma\gamma}^2$, 
where $\kappa \equiv \alpha \beta c_1/(2\pi f_\pi^2)$, 
$x_1 \equiv A_{n\pi^0}^2 + B_{n\pi^0}^2$, 
and $x_2 \equiv B_{n\pi^0}^2 - A_{n\pi^0}^2$.  The decay rate is then given by
\beq
\Gamma(n\to X \gamma\gamma) = \int_0^{(m_n - m_X)^2} ds_{\gamma\gamma}\, \frac{d \Gamma(n\to X \gamma\gamma)}{d s_{\gamma\gamma}}.  
\label{ratenX2a}
\eeq

For completeness, we will also give the rate for 
$n\to X \ell^+\ell^-$, via an off-shell $Z_d$, for $m_N-m_X < m_{Z_d}$, though 
this process is typically negligible.    
The differential decay rate, as a function of the $\ell^+\ell^-$ invariant mass $s_{\ell\ell}$ is given by
\begin{eqnarray}
&&\frac{d \Gamma(n\to X \ell^+\ell^-)}{d s_{\ell\ell}} = \frac{Q_d^2 \,\alpha_d \,\alpha \, \varepsilon^2}{24 \pi m_n^3}
\left(\frac{\beta \, c_1 \,m_X}{\mum^2}\right)^2 \label{diffnX2e}\\
&\times& \sqrt{(\mum^2 + s_{\ell\ell})^2 - 4 m_n^2 s_{\ell\ell}} \, 
\left[\frac{\mum^4 + s_{\ell\ell}(\mup^2 - 2s_{\ell\ell})}{(s_{\ell\ell} - m_{Z_d}^2)^2}\right] ,\nonumber
\end{eqnarray}
and the decay rate is 
obtained from  $\Gamma(n\to X \ell^+\ell^-) = 
\int_0^{(m_n - m_X)^2} ds_{\ell\ell}\, d \Gamma(n\to X \ell^+\ell^-)/d s_{\ell\ell}$.


\end{document}